\documentclass[preprint]{aastex} 

\shorttitle{A virtual X-ray of HAT-P-13b} 
\shortauthors{Batygin et al.} 

\newcommand{\be}{\begin{equation}}
\newcommand{\ee}{\end{equation}}

\def\lta{\,\raise 0.3 ex\hbox{$ < $}\kern -0.75 em
 \lower 0.7 ex\hbox{$\sim$}\,}
\def\gta{\,\raise 0.3 ex\hbox{$ > $}\kern -0.75 em
 \lower 0.7 ex\hbox{$\sim$}\,}

\begin{document}
 
\title{Determination of the Interior Structure of Transiting Planets in Multiple-Planet Systems}  

\author{Konstantin Batygin$^{1}$, Peter Bodenheimer$^{2}$, Gregory Laughlin$^{2}$ } 

\affil{$^1$Division of Geological and Planetary Sciences, California Institute of Technology, Pasadena, CA 91125} 
\affil{$^2$UCO/Lick Observatory, University of California, Santa Cruz, CA 95064}

\slugcomment{Submitted to: {\it Astrophysical Journal Letters}}
\email{kbatygin@gps.caltech.edu}
 
\begin{abstract} 

Tidal dissipation within a short-period transiting extrasolar planet perturbed by a companion object can drive orbital evolution of the system to a so-called tidal fixed point, in which the apsidal lines of the transiting planet and its perturber are aligned, and for which variations in the orbital eccentricities of both planet and perturber are damped out. Maintenance of the fixed-point apsidal alignment requires the orbits of both planet and perturber to precess at the same rate. Significant contributions to the apsidal precession rate are made by the secular planet-planet interaction, by general relativity, and by the gravitational quadropole fields created by the transiting planet's tidal and rotational distortions. The precession arising from the planetary quadrupole can be the dominant term, and is strongly dependent on the planet's internal density distribution, which is in turn controlled by the fractional mass of the planet incorporated 
into a heavy-element core. The fixed-point orbital eccentricity of the inner planet is therefore a strong function of the planet's interior structure.
We illustrate these ideas in the specific context of the recently discovered HAT-P-13 exo-planetary system, and
show that one can already glean important insights into the physical properties of the inner $P=2.91\, {\rm d}$, $M=0.85\, M_{\rm Jup}$, $R=1.28\, R_{\rm Jup}$ transiting planet.
We present structural models of the planet, which indicate that its observed radius can be maintained for a one-parameter sequence 
of models that properly vary core mass and tidal energy dissipation in the interior. We use an octopole-order secular theory of the orbital dynamics to derive the dependence of the inner planet's eccentricity, $e_b$, on its tidal Love number, ${k_2}_{b}$. We find that the currently measured eccentricity, $e_b=0.021\pm0.009$ implies $0.116<{k_2}_{b}<0.425$, $0 M_{\oplus}<M_{core}<120 M_{\oplus}$, and $Q_{ b}<300,000$.  Improved measurement of the eccentricity via transit and secondary eclipse timing, along with continued radial velocity monitoring, will soon allow for far tighter limits to be placed on all three of these quantities, and will provide an unprecedented probe into the interior structure of an extrasolar planet.
\end{abstract}


\section{Introduction} 

The mounting detection rate for extrasolar planets inevitably produces a series of ``firsts". Notable
examples include 51 Peg b, the first Jovian-mass planet orbiting a main-sequence star (Mayor \& Queloz 1995), Upsilon Andromedae
b, c, and d, the first multiple extrasolar planet system (Butler et al. 1999), HD 209458 b, the first transiting planet (Charbonneau et al. 2000,
Henry et al. 2000), and Gliese 581 e, the first truly terrestrial-mass planet (Mayor et al. 2009). 

The HAT-P-13 system (Bakos et al. 2009) presents an exoplanetary first that at first glance seems to lie perhaps one notch down on the novelty scale. This system contains the first transiting planet (HAT-P-13-b) that is accompanied by a well-characterized longer-period companion planet (HAT-P-13-c).

Transiting planets in multiple-planet systems have, however, been eagerly anticipated by the astronomical community. In configurations of this
type, the planet-planet perturbations can lead to a host of observational effects (largely involving precise timing of the transits) that will
potentially enable remarkable dynamical characterization of both the orbital and the planetary properties (see, for example, the
review of Fabrycky 2009). On this count, the HAT-P-13 system does not disappoint. We show here that a combination of a tidal-secular 
orbital evolution model, coupled with interior evolution models of the inner planet, can be used to probe the planet's interior structure and to measure its current tidal quality factor, $Q$. Using information currently available, we outline this approach, and show that
HAT-P-13-b has $Q<300,000$.

The plan of this paper is as follows. In section 2, we describe the dynamics of a system at a tidal fixed point, and we outline the resulting connection
between the interior structure of the planet and its orbital eccentricity. In section 3, we give an overview of the system and describe our interior evolution calculations. 
We show that, under the assumption of a tidal origin for the planet's inflated size, the observed planetary radius can be explained by a one-parameter sequence of models within a two-parameter space delineated by planetary core-mass and planetary tidal luminosity. We then summarize the prospects for improved measurement of the orbital eccentricity, and proselytize the overall ramifications of our study.

\section{A System at an Eccentricity Fixed Point}

In a planetary system that resides far from a significant mean motion resonance, the non-Keplerian portion of the orbital motion can be well described by secular terms in the planetary disturbing function. In this paper, we focus on the particular circumstance in which a short-period transiting planet on a nearly circular orbit receives perturbations from a relatively distant companion planet lying on a significantly eccentric orbit. Due to the high eccentricity of the outer body, classical Laplace-Lagrange theory (e.g. Murray \& Dermott 1999) cannot be used. Systems of this type are, however, well-described by  the octopole-order secular theory presented in Mardling (2007). This theory requires that (1) the eccentricity of the inner planet be much less than that of the outer planet, and (2) that the mass of the inner planet be much less than that of the star. The theory makes no strong demands, however, on either the mass of the outer planet or on its eccentricity. If we assume coplanar orbits, the secular evolution generated by the planet-planet interaction is given by:

\begin{equation}
\dot{e_b}=-\frac{15}{16} n_{b} e_{c}\left( \frac{m_{c}}{M_{\star}} \right) \left( \frac{a_{b}}{a_{c}} \right)^{4} \frac{\sin(\varpi_{b} - \varpi_{c})}{(1- e_{c}^{2})^{5/2}}-e_{b}\frac{21 \pi}{P_{b}} \frac{{k_{2}}_{b}}{Q_{b}} \frac{M_{\star}}{m_{b}} \left(\frac{R_{b}}{a_{b}}\right)^{5} \, ,
\end{equation}
\begin{equation}
\dot{e_c}=\frac{15}{16} n_{c} e_{b}\left( \frac{m_{b}}{M_{\star}} \right) \left( \frac{a_{b}}{a_{c}} \right)^{3} \frac{\sin(\varpi_{b} - \varpi_{c})}{(1- e_{c}^{2})^{2}} \, ,
\end{equation}
\begin{equation}
\dot{ \varpi_{b}}_{secular}=\frac{3}{4} n_{b} \left( \frac{m_{c}}{M_{\star}} \right) \left( \frac{a_{b}}{a_{c}} \right)^{3}\frac{1}{(1- e_{c}^{2})^{3/2}} \left[ 1- \frac{5}{4} \left( \frac{a_{b}}{a_{c}} \right) \left( \frac{e_{c}}{e_{b}} \right) \frac{\cos(\varpi_{b} - \varpi_{c})}{1- e_{c}^{2}} \right]\, ,
\end{equation}
and
\begin{equation}
\dot{ \varpi_{c}}_{secular}=\frac{3}{4} n_{c} \left( \frac{m_{b}}{M_{\star}} \right) \left( \frac{a_{b}}{a_{c}} \right)^{2} \frac{1}{(1- e_{c}^{2})^{2}} \left[ 1- \frac{5}{4} \left( \frac{a_{b}}{a_{c}} \right)\left( \frac{e_{b}}{e_{c}} \right) \frac{1+4e_{c}^2}{1-e_{c}^2}  \cos(\varpi_{b} - \varpi_{c}) \right] \, ,
\end{equation}
where the orbital elements take on their standard notation and specifically, ${k_2}_{b}$ is the tidal Love number, and $Q_b$ is the inner planet's effective tidal dissipation parameter. 

The inner planet experiences additional contributions to its precession from the quadrupole potential that arises from the tidal and rotational bulges of the planet, and from the leading-order effects of general relativity. As discussed in Raggozine \& Wolf (2009), precession driven by the tidal and rotational bulges of the star is unimportant, unless the rotational period of the star is short (e.g. $P_{rot} \lesssim 10$ days). Derivations of the planet-induced tidal and rotational precessions are given in Sterne (1939), and are discussed in the planetary context by Wu \& Goldreich (2002) and also, extensively, by Raggozine \& Wolf (2009).  The relativistic advance has been discussed by many authors, for an up-to-date discussion in the extrasolar planet context see, for example Jordan \& Bakos (2008). To linear order, we can treat the total precession of the inner planet as the sum of the four most significant contributions:
\begin{equation}
{\dot{\varpi_{b}}}_{total}={\dot{\varpi_{b}}}_{secular}+{\dot{\varpi_{b}}}_{tidal}+{\dot{\varpi_{b}}}_{GR}+{\dot{\varpi_{b}}}_{rotational} \, ,
\end{equation}
where, assuming synchronous rotation,
\begin{equation}
{\dot{\varpi_{b}}}_{tidal}=\frac{15}{2} {{k_{2}}_{b}}\left( \frac{R_{b}}{a_b} \right) ^{5}  \left( \frac{M_{b}}{M_{\star}} \right) f_{2}(e_{b}) n_b\, ,
\end{equation}
\begin{equation}
{\dot{\varpi_{b}}}_{GR}=\frac{3 n_{b}^{3}}{1-e_{b}^{2}} \left( \frac{a_{b}}{c} \right)^{2}\, ,
\end{equation}
and
\begin{equation}
{\dot{\varpi_{b}}}_{rotational}=\frac{{k_{2}}_{b}}{2} \left( \frac{R_{b}}{a_b} \right) ^{5} \frac{n_b^{3} a_b^{3}}{G m_{b}} \left( 1-e_b^{2} \right)^{-2} \, .
\end{equation}
The eccentricity function, $f_{2}(e_b)$ is given by
\begin{equation}
f_{2}(e_{b})=\left(1-e_{b}^{-2}\right)^{-5}(1+{3\over{2}}e_{b}^{2}+{1\over{8}}{e_{b}^{4}}) \, .
\end{equation}
Tidal dissipation occurs primarily within the inner planet, and leads to continual decrease of the inner planet's semi-major axis through
\begin{equation}
\dot{a_b}=-e_{b}^{2} a_{b} \frac{21 \pi}{2P_{b}} \frac{{k_{2}}_{b}}{Q_{b}} \frac{M_{\star}}{m_{b}} \left(\frac{R_{b}}{a_{b}}\right)^{5} \, .  
\end{equation}

When a system of the type modeled above is subjected to tidal friction, it evolves to a stationary configuration or a ``fixed point" within $\sim$ 3 circularization timescales (see Mardling 2007 for an in-depth discussion). Formally, a secular fixed point can be characterized by simultaneously aligned (or anti-aligned) apses and identical precession rates of the orbits. In other words, in the frame that precesses with the orbits, the system in stationary. It then follows (in the limit of large $Q_b$) that when ${\dot{\varpi_{b}}}_{total}={\dot{\varpi_{c}}}_{secular}$, we have $\dot{e_b}=\dot{e_c}=0$. When a fixed-point system is subjected to tidal dissipation (that is, has a finite $Q_b$) the eccentricities of both orbits decay slowly, and the system remains quasi-stationary. 

To second order in eccentricity, the tidal luminosity of a
spin-synchronous planet is given by (e.g. Peale \& Cassen 1978)
\begin{equation}
{dE_b\over{dt}}={21\over{2}}{{k_{2}}_{b}\over{Q_b}}{G M_{\star}^{2} n_b R_{b}^{5} e_b^{2}\over{a_b^{6}}} \, ,
\end{equation}
Note that if $e_b>0$ then the planet cannot be fully spin synchronized. Further, if the planet is a fluid body, it will be unable to maintain a permanent quadropole moment, and will therefore not reside in spin-orbit resonance. The pseudo-synchronization theory of Hut (1981); see also Goldreich \& Peale (1966) can be used to calculate the spin frequency (which for small $e_b$ approaches $n_b$)
\begin{equation}
{\Omega_{\rm spin}\over{n_b}}={1 + {15\over{2}}e_b^{2} + {45\over{8}}e_b^{4}
+{5\over{16}}e_b^{6} \over {(1 + 3e_b^{2} + {3\over{8}}e_b^{4})(1-e_b^{2})^{3/2}} }\,
\end{equation}
The analysis of Levrard et al. (2007), furthermore, indicates that this spin asynchronicity of the planet will cause the tidal luminosity to exceed that given
by the above formula by a small amount. 

The tidal Love number, ${k_{2}}_{b}$ parameterizes the degree of central condensation in the fluid inner transiting planet. The mass distribution in turn
affects the total tidal luminosity through Equation (11) and contributes to the orbital precession rate through ${\dot{\varpi_{b}}}_{tidal}$. The quantity ${k_{2}}_{b}$ therefore
provides an explicit connection between the interior sturcture and energetics of the planet on the one hand, and the orbital
dynamics on the other. If the density distribution, $\rho(r)$, in a planet is available, then calculation of ${k_2}_{b}$ is straightforward (Sterne 1939):\footnote{note that the quantity $k_{2,1}$ defined in Sterne (1939) is the {\it apsidal motion constant}, that is, ${{k_2}_b}/2$ in our notation}
\begin{equation}
{k_2}_{b}={3-\eta_2(R_{Pl})\over{2+\eta_2(R_{Pl})}}\, ,
\end{equation}
where $\eta_2(R_{Pl})$ is obtained by integrating an ordinary differential equation for $\eta_{2}(r)$ radially outward from $\eta_2(0)=0$.
\begin{equation}
r{d\eta_{2}\over{dr}}+\eta_{2}^{2}-\eta_{2}-6+{6\rho\over{\rho_m}}(\eta_2+1)=0\, ,
\end{equation}
where $\rho_m$ is the mean density
interior to $r$ and $R_{\rm Pl}$ is the outer radius of the planet.

\section{Application to the HAT-P-13 Planetary System}

The theory discussed above finds an ideal application in the context of HAT-P-13. As discussed in Bakos et al (2009, hereafter B09) this system
was discovered in a wide-field photometric survey, and was later confirmed and characterized with high-precision radial velocities.  The system contains an inner, transiting, jovian-mass planet, ``b", and an outer body, ``c", with $M_{c}\sin(i_c)$ close to the giant-planet brown dwarf boundary.
The $V=10.65$ G4V parent star, formerly best known as GSC 3416-00543, was essentially unstudied prior to the photometric detection of its
inner planet, and so, as a result, all quoted planetary and stellar properties are drawn from the B09 discovery paper.
For reference, we note that B09 derive an inner planet mass of $m_{b}=0.851^{+0.029}_{-0.046} \,M_{\rm Jup}$, a period $P_b=2.91626 \pm 0.00001$ days, and an eccentricity (measured from a fit to 32 radial velocity measurements) of $e_b=0.021\pm0.009$.
The outer companion has $M_{c}\sin(i_c)=15.2\pm1.0 \,M_{\rm Jup}$, period $P_c=428.5 \pm 3.0$ days, and eccentricity $e_c=0.691\pm0.018$. To within the significant observational uncertainty, the apsidal lines of the two planets are aligned. The
parent star has $M_{\star}=1.22^{+0.05}_{-0.10}\, M_{\odot}$, and the planetary radius is $R=1.28\pm0.079 R_{\rm Jup}$.

The short orbital period and non-zero eccentricity of HAT-P-13b suggest that tidal circularization should be highly effective over
the presumed multi-Gyr age of the star. In this case, dissipative secular evolution will have brought the system to a fixed point configuration, which to high accuracy satisfies the constraint given by 
\begin{equation}
{\dot{\varpi_{b}}}_{total} (e_b , {k_2}_b)={\dot{\varpi_{c}}}_{secular}\, .
\end{equation}
We note that the system as currently characterized from the observations is fully consistent with such a configuration.

Our approach is illustrated schematically in Figure 1, and operates as follows.
For given $M_{\rm Pl}$, $T_{\rm eff}$ and $R_{\rm Pl}$
(all of which are strongly constrained by the observations) we compute planetary interior structure and evolution models with a descendant
of the Berkeley stellar evolution code (Henyey et al. 1964). This program assumes that the
standard equations of stellar structure apply, and it has been used extensively in the study of both extrasolar
and solar system giant planets (see e.g.  Pollack et al. 1996,
Bodenheimer et al. 2003, Hubickyj et al. 2005, and Dodson-Robinson
et al. 2008 for descriptions of the method and its input physics). 
Energy sources within the planet include gravitational contraction,
cooling of the interior, and tidal heating of the interior, which we
assume occurs at adiabatic depth.
At the planetary surface, the luminosity is composed of two components:
the internal luminosity generated by the planet, $L_{int}=dE/dt$, and 
the energy absorbed from the stellar radiation flux 
and re-radiated (ÔinsolationÕ). Pure molecular opacities are used in the
radiative outer layers of the planet (Freedman et al. 2008).
A given evolutionary sequence starts at a 
radius of roughly 2 $R_{\rm Jup}$ and ends at an age of 4.5 Gyr. The
resulting planetary radius is highly insensitive to the chosen age. It is
assumed that the planet arrived at its present orbital position during
or shortly after formation, that is, at an age of $< 10^7$ yr.

We used the code to delineate a range of plausible models for HAT-P-13b,
all of which are required to match the observed planetary mass,
$M=0.85 M_{\rm Jup}$, and the inferred planetary effective temperature,
$T_{\rm eff}=1649\, {\rm K}$. Our models are divided into three sequences,
 which are required to match the observed one-sigma lower limit on the
planetary radius, $R_{\rm {Pl}}=1.20 \, R_{\rm Jup}$ (sequence 1), the best-fit
radius $R_{\rm {Pl}}=1.28 \, R_{\rm Jup}$ (sequence 2), and the one-sigma upper
 limit on the observed radius, $R_{\rm {Pl}}=1.36 \, R_{\rm Jup}$ (sequence 3).
The structurally relevant unknown parameters are the planet's solid core
mass, $M_{core}$, and the total tidal luminosity, $L_{int}=dE/dt$. By computing a variety of models
in this 2-dimensional parameter space, we can  pin down
the ($M_{\rm core}$, $L_{int}$) pairs that generate planets that satisfy a
given choice of ($M_{\rm Pl}$, $R_{\rm Pl}$, $T_{\rm eff}$). 
A range-spanning aggregate of the models is listed in Table 1. 

Also listed in Table 1 are the tidal Love numbers, ${k_2}_{b}$ (obtained from $\rho(r)$ via Equations 13 and 14), the tidal quality factors, $Q_{b}$ (obtained from equation 11), and the fixed point eccentricities, $e_b$ that enable the satisfaction of equation 15. Quoted errors on $Q_b$ and $e_b$ are obtained by using B09's reported uncertainties on the observed planetary and orbital properties, and adopting the assumptions that the error distributions are both normally distributed and uncorrelated across parameters.

Given our aggregate of models, we can consider the effect of the Love number, ${k_{2}}_{b}$ on the orbital architecture in more detail. If we completely ignore the orbital precession induced by the planet's tidal and rotational bulges, and adopt B09's best-fit measurements of all relevant orbital parameters other than $e_b$, the equilibrium inner planet eccentricity is $e_{b}^{eq} = 0.0336$. We verified with a fully numerical 3-body simulation (Chambers 1999), that includes the relativistic precession, that this value is indeed an excellent approximation to the fixed-point eccentricity. If we assume a Love number, ${k_{2}}_{b}= 0.3$, and include the precession due to planetary rotational and tidal bulges, the equilibrium eccentricity drops to $e_{b}^{eq} = 0.0161$, a difference, $\Delta e_b$, that is eminently detectable. Therefore, with the precise estimate of planet b's eccentricity, that will emerge from secondary transit timing and additional radial velocity measurements, it will be possible to make solid inferences about planet b's core mass and internal luminosity from direct measurement of the tidal Love number,  ${k_{2}}_{b}$. In the context of this system, it is important to note that the orbital precession of the planets is quite slow ($\sim 10^{-3}$ deg/yr). As a result, ${k_{2}}_{b}$ cannot be measured directly from transit light curves, as described in (Ragozzine \& Wolf, 2009), and must be inferred from the equilibrium eccentricity.

To further illustrate this idea, we obtained a series of equilibrium eccentricity values as a function of ${k_{2}}_{b}$. The results from these calculations are shown in figure 2. Modeling the errors as described above, the mean trend of planet b's eccentricity can be approximated by a fourth-order polynomial as
\begin{equation}
e_{b}^{eq} \approx 0.0334 - 0.0985 {k_{2}}_b +0.188 {k_{2}}_b^{2} - 0.184 {k_{2}}_{b}^{3} + 0.069 {k_{2}}_{b}^{4}.
\end{equation}
The plotted errors of the equilibrium eccentricities are the standard deviations obtained from each sample of $e_{b}^{eq}$'s for a given ${k_{2}}_{b}$. We stress that these error bars will shrink very significantly with improved observational measurements obtainable from photometry, timing, and radial velocity.

In addition to the mean trend and errors, specific regions in ($e_b,{k_{2}}_{b}$) space, occupied by a set of  interior models with 0, 40, 80, and 120 $M_{\oplus}$ core masses (see Table 1) are also marked. These regions are represented as four quadrilaterals overlaying the graph. The corners of each quadrilateral correspond to the combination of the Love number for a given model, specific to a radius of 1.2 $R_{Jup}$ or 1.36 $R_{Jup}$, and 1-$\sigma$ bounds on its equilibrium fixed point eccentricity as determined by Equation 15. An increased core mass tends to lower the Love number. Accordingly, the left-most quadrilateral on Figure 2 corresponds to the 120 $M_{\oplus}$ core model, while the right-most quadrilateral represents the core-less model. 

Recall again, that we have neglected the precession induced by the star's tidal bulge and rotation. While unlikely, if these effects shall turn out to be important, they will lead to a further decrease in the equilibrium eccentricity of planet b. If the star rotates rapidly, a degeneracy will appear with respect to probing the interior structure of planet b, as the star's unknown Love number, ${k_{2}}_{\star}$, will enter the calculation. In addition, we are assuming a co-planar configuration. The validity of this assumption will be tested by forthcoming transit timing measurements, and a further clue will be provided by measurement of the alignment of the inner planet's orbit with the stellar equator via the Rossiter-McLaughlin effect.

Given the non-zero eccentricity of planet b, a natural question emerges: how long will planet b remain slightly eccentric despite tidal dissipation? To answer this question, we performed a tidally dissipated secular integration, similar to those discussed in Batygin et al. (2009) using an artificially low tidal quality factor of $Q_b=10$ to speed up the proceedings. This integration revealed that the {\it e}-folding time for planet b's eccentricity is $\tau \approx  5.78 (Q_b/10) \times 10^{5}$ years, or approximately 6 Gyr for $Q_b=10^{5}$. As a consequence, we expect that the orbital configuration of the system {\it has} evolved somewhat during the current lifetime of the star. This timescale also places the low values of $Q_b$, e.g. $Q_{b}<10,000$, that are currently admitted by the observations into disfavor.

\section{Conclusion}

Our analysis indicates that the HAT-P-13 system has the near-immediate potential to give startlingly detailed information about the density structure and the efficiency of tidal dissipation in the interior of an extrasolar planet. When high-precision (yet fully feasible) refinements of the orbital parameters are obtained, we will gain a precise and accurate measurement of the tidal quality factor, $Q_b$, of HAT-P-13b -- superior, in fact, to those that we currently have for the solar system giant planets. Furthermore, it is seems reasonable to assume that additional examples of systems that contain a transiting planet at a well characterized tidal fixed point will soon emerge from the ongoing photometric and doppler velocity surveys.

We therefore encourage immediate observational effort to obtain an improved characterization of the HAT-P-13 system, and we reiterate
the importance of the wide-field surveys (such as HAT Net) that can locate transiting planets orbiting the {\it brightest available} stars in the sky.

This research is based in part upon work supported by NASA Grant NNX08AH82G
(PB) and by the National Science Foundation CAREER program under Grant No. 0449986
(GL).

\newpage 
%\vskip1.0truein 

\clearpage
\begin{table}
\begin{center}
\caption{Stationary Orbital and Interior Models of HAT-P-13b }
\begin{tabular}{cccccccccc}
\tableline\tableline
Core Mass ($M_{\oplus}$) & R ($R_{Jup}$)  & dE/dt ($erg/s$) & $k_{2}$ & Q & e &\\
\tableline
\\
0 & 1.20 & 3.49 $\times 10^{25}$ & 0.425 & 228561 $\pm$ 65014 & 0.0165 $\pm$ 0.0032  &\\
0 & 1.28 & 1.78 $\times 10^{26}$ & 0.38 & 42054 $\pm$ 14864 & 0.0143 $\pm$ 0.0036  &\\
0 & 1.36 & 5.6 $\times 10^{26}$ & 0.34 & 12875 $\pm$ 4396 & 0.0129 $\pm$ 0.0030  &\\
\\
40 & 1.20 & 2.485 $\times 10^{26}$ & 0.297 & 30868 $\pm$ 8603 &  0.0195 $\pm$ 0.0036  &\\
40 & 1.28 & 8.45 $\times 10^{26}$ & 0.26 & 8924$\pm$ 2798 & 0.0176 $\pm$ 0.0037  &\\
40 & 1.36 & 1.93 $\times 10^{27}$ & 0.228 & 3959 $\pm$ 1154 & 0.0162 $\pm$ 0.0034  &\\
\\
80 & 1.20 & 1.028 $\times 10^{27}$ & 0.217 & 6810 $\pm$ 2049 & 0.0221 $\pm$ 0.0049  &\\
80 & 1.28 & 2.53 $\times 10^{27}$ & 0.187 & 3036 $\pm$ 929 & 0.021 $\pm$ 0.0041  &\\
80 & 1.36 & 4.96 $\times 10^{27}$ & 0.163 & 1535 $\pm$ 453 & 0.0193 $\pm$ 0.0038  &\\
\\
120 & 1.20 & 3.25 $\times 10^{27}$ & 0.159 & 1967 $\pm$ 618 & 0.0246 $\pm$ 0.0046  &\\
120 & 1.28 & 6.88 $\times 10^{27}$ & 0.135 & 997 $\pm$ 273 & 0.0226 $\pm$ 0.0039  &\\
120 & 1.36 & 1.27 $\times 10^{28}$ & 0.116 & 563 $\pm$ 182 & 0.0216 $\pm$ 0.0042  &\\
\\

\tableline
\end{tabular}
\end{center}
\end{table}

\clearpage

\begin{figure*}[t]
\centering
\includegraphics[width=1.0\textwidth]{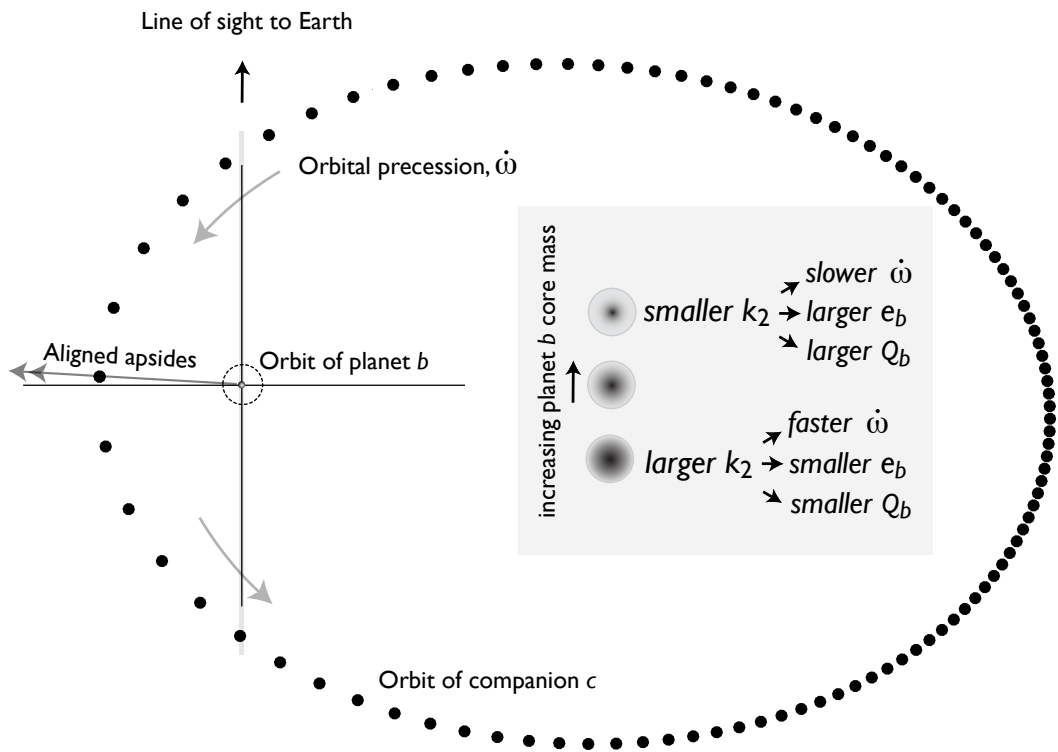}
\caption{A representation of the orbital architecture of the HAT-P-13 system to scale. The inset schematic illustrates the dependencies of ${k_2}_{b}$, $Q_b$, $\dot{\varpi}$ and $e_b$ on the mass of the planet's heavy element core.}
\end{figure*}

\begin{figure*}[t]
\centering
\includegraphics[width=1.0\textwidth]{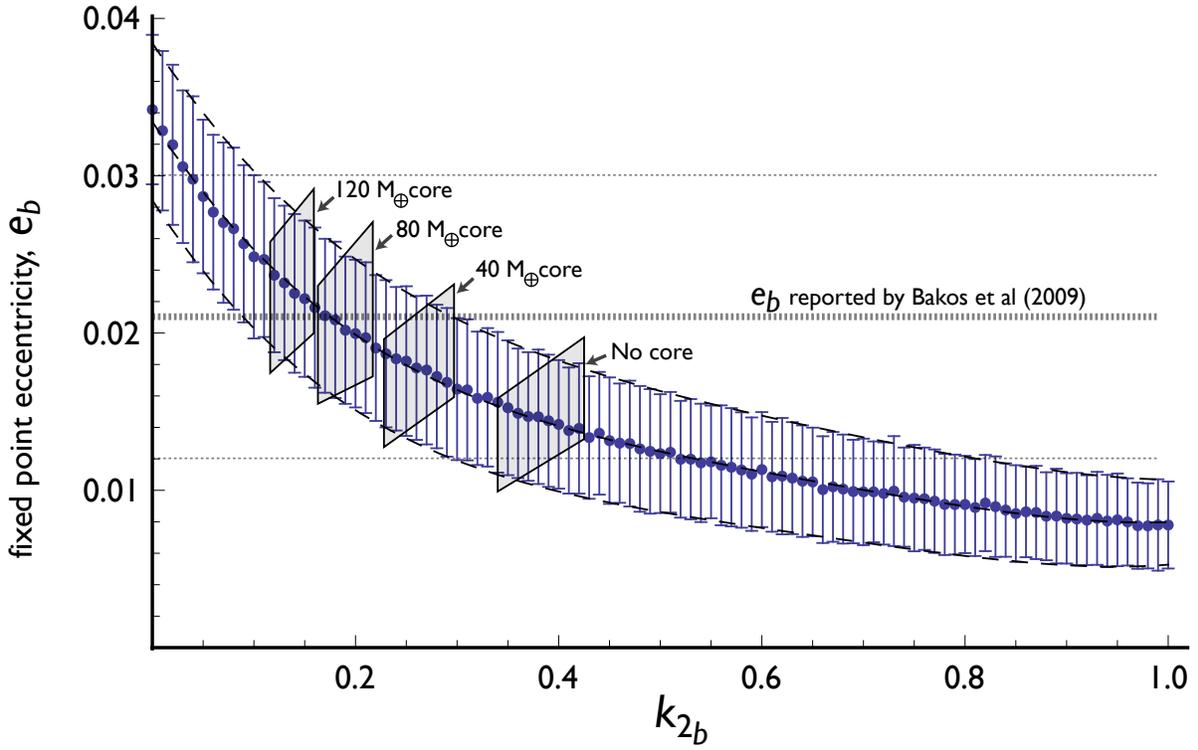}
\caption{Stationary eccentricity of HAT-P-13b as a function of its Love number, ${k_2}_{b}$, with error bars. Each blue dot represents the sample mean of the computed fixed point eccentricities. The dashed lines are best-fit fourth order polynomials. The four quadrilaterals are the approximate regions of the ($e_b,{k_{2}}_{b}$) space occupied by each of the models presented in Table 1.}
\end{figure*}

\end{document}